\documentclass{ws-ijmpd}
\usepackage{epsfig}
\usepackage{epstopdf}
\usepackage{color}

\newcommand*\DAlambert{\mathop{}\!\mathbin\Box}

\begin{document}

\title{Gravitational wave: generation and detection techniques}

\author{Saibal Ray}
\address{Centre for Cosmology, Astrophysics and Space Science (CCASS), GLA University, Mathura 281406, Uttar Pradesh, India\\saibal.ray@gla.ac.in}

\author{R. Bhattacharya}
\address{Department of Physics and Astronomy, University of New Mexico, Albuquerque, New Mexico, USA\\rbhattacharya1995@unm.edu}

\author{Sanjay K. Sahay}
\address{Department of Computer Sciences and Information Systems, BITS, Pilani, K.K. Birla Goa Campus NH-17B, By Pass Road Zuarinagar 403726, Goa, India\\ssahay@goa.bits-pilani.ac.in}

\author{Abdul Aziz}
\address{Department of Physics, Bodai High School (H.S.), Bodai, Amdanga, North 24 Parganas 700126, West Bengal, India\\azizmail2012@gmail.com}

\author{Amit Das}
\address{Department of Physics, Ashoknagar Vidyasagar Bani Bhaban High School (H.S.), Ashoknagar, North 24 Parganas 743222, West Bengal, India\\amdphy@gmail.com}

\maketitle	

\abstract{In this paper, we review the theoretical basis for generation of gravitational waves and the detection techniques used to detect a gravitational wave. To materialize this goal in a thorough way we first start with a mathematical background for general relativity from which a clue for gravitational wave was conceived by Einstein. Thereafter we give the classification scheme of gravitational waves such as (i) continuous gravitational waves, (ii) compact binary inspiral gravitational waves and (iii) stochastic gravitational wave. Necessary mathematical insight into gravitational waves from binaries are also dealt with which follows detection of gravitational waves based on the frequency classification. Ground based observatories as well as space borne gravitational wave detectors are discussed in a length. We have provided an overview on the inflationary gravitational waves. In connection to data analysis by matched filtering there are a few highlights on the techniques, e.g. (i) Random noise, (ii) power spectrum, (iii)  shot noise, and (iv) Gaussian noise. Optimal detection statistics for a gravitational wave detection is also in the pipeline of the discussion along with detailed necessity of the matched filter and deep learning.}

\keywords{general relativity; gravitational waves; detection techniques, LIGO}

\section{Introduction}
 The first sign of gravitational wave (GW) was conceived theoretically by Albert Einstein in the year 1916 in two consecutive papers under general relativity (GR) with a trial and error method and later on confirmed his results in the year 1918 by analysing and correcting mistake which occured in the previous papers~\cite{Einstein1916a,Einstein1916b,Einstein1918,Weinstein2016,Dirkes2018}. Einstein could predict that though the nature of GWs are similar to that of electromagnetic radiations but, unlike electromagnetism where dipole source emits waves, he considered quadrupole source to describe the rate of energy decay due to emission of spacetime ripples from a binary mechanical system~\cite{Einstein1918}. 

\begin{figure}[h]
    \centering
    \includegraphics[width=6cm]{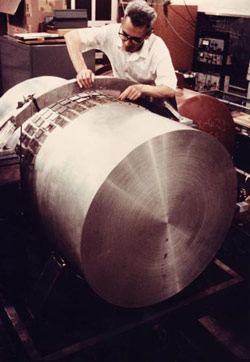}
    \caption{Weber is adjusting the instrumentation: one of his aluminum cylinders, 2 meters in length and 1 meter in diameter, for detecting gravitational waves. [Image Credit: https://physics.aps.org/story/v16/st19]}
    \label{fig1}
\end{figure}

As far as history of science is concerned Joseph Weber may be credited with conducting early research into the gravitational wave detection which began as early as in 1958 and continued till his later period of investigations. Though his reported results could not be duplicated by scientific experiments and created much debates among the experimental scientists but his work would inspire other scientists who have been working in the field. Today Weber is considered one of the pioneers in the front-line of spacetime ripple, i.e. GW detection, especially via the {\it Weber Bar} -- the first GW detector (Fig. \ref{fig1})~\cite{Weber1958}.

 In seventies the first binary neutron star system was discovered which served as a proof for the prediction of GR that binary system generally evolve with time which supports the idea of evolution of a moving mass system due to energy dissipation. Thus it was conceptualised that gravitational wave is emitted by all kinds of binary system. In this connection we would like to mention the experimental evidences on GW detection which came from the binary pulsar PSR B1913+16 (or popularly known as the {\it Hulse-Taylor binary pulsar}). This was first discovered in 1974 by Joseph Taylor (Jr.) and Russell Hulse and interestingly after two decades that helped them to win the 1993 Nobel Prize in Physics. It has been noticed that the newly discovered pulsar PSR B1913+16 was exhibiting a regularly varied pulsating rate. Thus the discovery is basically a powerful evidence that proves Einstein’s General Relativistic prediction of energy decay. Interestingly, the ratio of the observed to predicted rate of orbital energy loss is calculated to be 0.997±0.002 and hence remarkably in coherence to the prediction of GR~\cite{Taylor1981,Taylor1982,Taylor1989}.

Recent researches also show that apart from black hole or neutron stars, binary white dwarfs may also give rise to GW~\cite{Oliveira2020}. Every binary system emitting  GW experiences change in their orbital frequencies. The inevitability of this fact is due to the energy dissipation mechanism which explains the reduction in orbital energy and the emission of GW at the expense of it as conceived by Einstein (mathematical approach in calculating this for a binary system is given in later Section). However, it is to be noted that for each binary system emitting GW there are roughly three stages; inspiral, coalescence and merger. The amplitude of the gravitational ripple signal obtained in case of binary systems is different in each stage. Thus only for a limited amount of time, detectors can detect gravitational ripples from a binary black hole system.
 
However, in recent times, specifically the last decade has been very much productive for GW astronomy through ground breaking experimental evidences. In 2015 LIGO detected the first spacetime ripple from a binary black hole GW150914 which was obviously a huge breakthrough in this field and paved the way to do new researches in fields of Gravitational wave astronomy~\cite{Abbott1,Abbott2}. These signals were observed by the LIGO's twin observatories, the LIGO Scientific Collaboration and Virgo Collaboration, on September 14, 2015 and later on February 11, 2016 announced the first confirmed observation. Therefore, quite interesting coinciding fact is that nearly 100 years after Einstein’s prediction of GR-based gravitational ripples it was a direct detection of the enigmatic idea. 

Regarding very recent status of gravitational wave astronomical researches on some other related aspects we would specifically like to mention various pulsar timing array (PTA) experiments (such as NANOGrav, EPTA, PPTA, CPTA, including data from InPTA). PTA collaborations have independently reported intersting evidence for a stochastic gravitational-wave background (SGWB), which can unveil several cosmological issues, e.g. (i) magnetars may have relationship with SGWB and thus can be considered a potential candidate as gravitational wave sources, (ii) the formation of primordial seeds of inhomogeneities in the early universe, (iii) prediction of the primordial black hole (PBH) clusters from the domain walls of axion-like particles (ALPs), (iv) early active galactic nuclei (AGN) formation. There are sufficient indications and evidences in favour of such scenarion due to James Webb Space Telescope (JWST)~\cite{SRC2021,Guo2023}. It is worthy to note that Vagnozzi~\cite{Vagnozzi2023} has provided an inflationary interpretation of the SGWB signal detected by PTA experiments.  On the other hand, it have been argued by several scientists regarding possible primordial origin of coalescenting BHs in LIGO-VIRGO-KAGRA events, as well as possible relationship between GW and PBH~\cite{Abbott3,Abbott4}. In particular, PBH origin of the GW event has been well discussed in the following refs.~\cite{Khlopov2010,Belotsky2019}. 

In this review, to have a deeper understanding of the spacetime ripple detection, we mainly focus on the theoretical aspects and the mathematical background of the  GW, their classification and also review certain aspects about their detection. Just to give a sense about GW detection we may compare this with other astronomical observatories. This leads to the fact that gravitational wave observatories are meant to feel the presence of GW signal rather than seeing it. In this review we also discuss about various types of observatories and the frequency range that are or would be used for their detection. 

Therefore, in the next few Sections as well as Subsections we discuss in detail the mathematical aspects of gravitational wave starting from GR intending slowly to build the theoretical foundations of GW. The overall scheme of the review presentation is as follows: a mathematical background for general relativity has been provided in Section 2. The classifications of GWs such as (i) continuous gravitational wave, (ii) compact binary inspiral gravitational wave and (iii) stochastic gravitational wave are shown in Section and Subsections 3. In Section 4, a few necessary mathematical steps into gravitational waves from binaries are dealt with which follows detection of GW based on the frequency classification. In the next Section 5, ground based observatories as well as space borne gravitational wave detectors are discussed in a length. We have provided an overview on the inflationary gravitational waves in Section 6. However, regarding processing of data by matched filtering, we have given a few highlights on the techniques, e.g. (i) Random noise, (ii) power spectrum, (iii)  shot noise, and (iv) Gaussian noise in Section 7. Regarding detection statistics for a gravitational wave we have provided respectively, in Sections 8 and 9, some recent techniques which specifically include the matched filter and deep learning. In Section 10, we put forward a few concluding remarks.

\section{Mathematical background of general relativity in connection to GW}

We begin with the following metric which gives the curved spacetime distance between two events in a four dimensional spacetime:
\begin{equation}
ds^{2}=g_{\mu\nu}dx^{\mu}dx^{\nu}.
\end{equation}

 The symmetric tensor $g_{\mu\nu}$ denotes the entire curvature of spacetime. The basic postulate of general relativity is that any presence of mass always bends the fabric of spacetime. Thus the curvature created can be quantified by the Riemannian tensor ($R_{\mu\nu\lambda\psi})$ which is a function of metric tensor, its first and second derivatives. Thus a flat space necessarily results in a vanished Riemannian tensor.
 
Eventually, the relation between the spacetime and matter can be provided as
\begin{equation}
 G_{\mu \nu}=\frac{8 \pi G}{c^4} T_{\mu \nu}
\end{equation}
where $G_{\mu \nu}$ is known as the Einstein tensor representing the curved spacetime geometry and $T_{\mu \nu}$ represents the energy momentum tensor. 

The Einstein tensor, again, can be expressed in the following way
 \begin{equation}
G_{\mu \nu}= R_{\mu \nu} - \frac{1}{2}R g_{\mu \nu}.
\end{equation}

The right hand side of the equation can be equated to $T_{\mu\nu}$ which is known as the stress energy tensor. Here $R_{\mu \nu}$ is known as Ricci tensor which is simply a contracted form of the Riemannian whereas $R$ is known as the Ricci scalar or scalar curvature which again is obtained by a contraction of Ricci tensor. Again, $ G_{\mu \nu}$ is known as Einstein tensor and G is the gravitational constant. A Ricci tensor vanishes when the spacetime is free from any distribution of matter. In case of absence of a Ricci tensor any expression for gravitational wave can be given by a Riemannian tensor.

In general relativity, gravity has a definite geometric interpretation and is obviously different from that in Newtonian description of gravity. In case of a four dimensional spacetime, a geodesic is nothing but the shortest curved distance between two points and for the curved space time geodesics must be calculated by taking the curvature into account. Moreover we can imagine that motion of a mass in gravitational field as moving along a geodesic. The given equation is known as a geodesic equation, where $\Gamma_{\lambda \mu}$ is the Christoffel’s symbol~\cite{Weinberg1972}.
 
Now the Christoffel’s symbol is given by~\cite{Weinberg1972,Carroll2004}
\begin{equation}
\Gamma^{\lambda}_{\mu\nu}=\frac{1}{2}\left[\frac {\partial g_{\lambda \beta}}{\partial x^{\nu}}+\frac {\partial g_{\mu\beta}}{\partial x^{\mu}}-\frac {\partial g_{\mu \nu}}{\partial x^{\beta}}\right].
\end{equation}

Since, the indices on each side can take values 0, 1, 2, 3. So, this whole set would comprise of 16 equations, thereby expressing the curvature of spacetime and this effect might be perceived as the gravitational wave. Here the index 0 corresponds to time and the index 1 corresponds to a particular direction in space, similarly by changing the indices, we can ascertain how the expression behaves in case of other dimensions.

For the vacuum case the energy momentum tensor reduces to
\begin{equation}
T_{\mu \nu}=\frac {\partial \Gamma^{\lambda}_{\mu\lambda}}{\partial x^{\nu}}+\frac {\partial \Gamma^{\lambda}_{\mu\nu}}{\partial x^{\lambda}}+\Gamma^{\beta}_{\mu\lambda}\Gamma^{\lambda}_{\nu\beta}-\Gamma^{\beta}_{\mu\nu}\Gamma^{\lambda}_{\beta\lambda}=0.
\end{equation}

A change in the stress-energy tensor indicates a changed matter distribution, which in turn gives rise to a modified gravitational field. Thus the metric gets perturbed and the change in this metric can be shown to be
\begin{equation}
\bar{g}_{\mu\nu}=g_{\mu\nu}+h_{\mu\nu}.
\end{equation}
 
In the above expression, $h_{\mu\nu}$ is the variation induced in spacetime metric. Basically, this new tensor can be regarded as the signature of the GW. This tensor can be calculated with the help of Einstein's field equation, which might be difficult to solve. Therefore we make use of the linearized theory of gravity, which assumes the perturbation to be very small ($h_{\mu\nu}<<1$) and only deals with the terms linear in $h_{\mu\nu}$.

Particularly, in the case of weak fields, this approximation gives correct result for generation and propagation of GW. Einstein formulated that GW propagates with a velocity equal to that of light. Now according to Einstein, linearized perturbation around a flat metric which will mathematically implicate $g_{\mu\nu}=\eta_{\mu\nu}$. The tensor will be given by
\begin{equation}
\bar{h}_{\mu\nu}=h_{\mu\nu}-\frac{1}{2}\eta_{\mu\nu}h_{\alpha}^{\alpha}.
\end{equation}

This tensor will obey the wave equation similar to electromagnetic wave equation, which accepts the plane wave solution. In the above equation $\bar{h}_{\mu\nu}$ is the gravitational field and $h_{\mu\nu}$ is the metric perturbation.

The wave equation thus can be written as
\begin{equation}
\DAlambert {\bar{h}_{\mu\nu}}=0,
\end{equation}
where $\DAlambert$ is known as D'Alembertian operator. The above equation is basically the three dimensional wave equation constrained by the condition known as Hilbert's gauge condition which is given by $\delta_{\mu}\bar{h}_{\mu\nu}=0$.

The gauge condition of Hilbert is similar to the Lorentz gauge condition used in electromagnetism. To begin with the gauge transformation there must be a suitable coordinate transformation. This is given by
\begin{equation}
x'^{\mu}=x^{\mu}+\varepsilon^{\mu},
\end{equation}
where $\varepsilon^{\mu}$ must satisfy the condition,
$\delta_{\mu}\delta^{\mu} \varepsilon^{\mu}=0$. This is essential, as if this condition does not satisfy, then the new gravitational field would not agree with the Hilbert gauge condition.

The gravitational wave equation has the simplest solution in
\begin{equation}
  \bar{h}_{\mu\nu}=A^{\mu\nu}e^{ik_{\alpha}x^{\alpha}}.  
\end{equation}

 When compared to generic wave equations, it can be inferred that $A^{\mu\nu}$ represents the amplitude and polarization of the wave, whereas, $k_{\alpha}$ gives the propagation direction and the frequency of the wave. Hilbert Gauge condition requires $A^{\mu\nu}k_{\alpha}=0$ in this case. The Hilbert Gauge condition also implies a geometrical meaning as the relation clearly shows the orthogonality of the two quantities. However, we are concerned only with the real portion of the solution.

\section{Classification of gravitational waves}
Gravitational waves may be characterized into several types based on their origin. Broadly, they can be classified into four types. We will review each type briefly.
\\
\\
1) \textbf{Continuous gravitational waves:}
Gravitational waves that are thought to be produced by a single spinning massive object, like a neutron star can be regarded as continuous GW. These signals from a neutron star can be attributed to the physical asymmetries in the star. Estimates shows that the continuous GW from neutron stars are very weak, approximately in the order of $10^{-25}$ or lesser. In general, the characteristic amplitude of GW from neutron stars can be given by the relation, $h(If_0^2\epsilon)/r$, where $f_0$ is the frequency of the emitted signal, $I$ is the moment of inertia of the neutron star, $\epsilon$ is the ellipticity denoting the asymmetric nature of the star and $r$ is the distance of the neutron star from the detector~\cite{Piccini}.
\\
\\
2) \textbf{Compact binary inspiral gravitational waves:} Binary objects, e.g. binary black hole or binary neutron stars, are one of the most prominent source of GW. The constituent objects of a binary system merges with each other, thereby generating GW in the procedure after spiralling towards each other for a long period of time~\cite{satyaprakash}.   
\\
\\
3) \textbf{Stochastic gravitational wave:} Apart from compact binary GW and continuous GW which are generally detected as a single event having higher intensity, there may be weaker gravitational wave signals coming from all over the cosmos at random times thereby giving rise to a GW background. This is generally referred to as stochastic GW signals. These types of signals are devoid of a particular waveform and hence can be tough to detect. The word stochastic basically denotes the fact that these signals have random waveform~\cite{Lawrence}. 
\\
\\
4) \textbf{Burst gravitational waves:}
If a star collapse non-spherically, then a GW signal might be emitted for a very small period of time (in the order of few milli-seconds). However, as modelling the collapse dynamics is a substantial challenge, burst GW signal waveform are very difficult to detect. However, one possible way of detecting the burst gravitational wave is by analyzing the excess power in the interferometric data. It has been shown~\cite{Ando} that for the peak power and the corresponding recorded time a threshold of $h = 6.6 \times 10^{-20}/(\sqrt{Hz})$ can be attained, which corresponds to $h = 3 \times 10^{-17}$ in strain for 1 ms short-burst signal with a 500 Hz analysis bandwidth. Given the order of the strain this type of signals can only be detected by LIGO/VIRGO, if the events take place within our galaxy. The probability of detecting this type of signals from our galaxy is very low considering the fact that rate of formation of supernovae is very less which is only one event in 50 years range.

\section{Mathematical insight into gravitational waves from binaries}

 Binary sources were the first source from which spacetime ripples were detected and historically, the first signal was detected by LIGO from a binary black hole, i.e. GW150914. In this Section we discuss some basic mathematical insights into detection of GW from binaries. To start with, let us assume the GW amplitude or strain to be $h$. As a direct consequence of conservation of energy $h$ is inversely proportional to the luminosity distance $r$~\cite{Bejger}.

 For a binary system let us consider the masses of two  bodies as $m_{1}$ and $m_{2}$ with a semi-major axis $a$. The total mass of the system is therefore taken to be 
 \begin{equation}
  M=m_{1}+m_{2}.  
 \end{equation}
 
 As we deal with binary system we must use the reduced mass $\mu=m_{1} m_{2}/M$. The quadrupole moment is given by
 \begin{equation}
  Q \propto \mu a^2.
 \end{equation}
 
 Again the strain amplitude $h$ is proportional to second derivative of the quadrupole moment. Thus $h$ can be given by
\begin{equation}
h \propto \frac{\ddot{Q}}{r}.
\end{equation}

The above equation basically represents the acceleration of the masses. Putting the value for quadrupole moment in the above equation we get
 \begin{equation}
h \varpropto \frac{\mu a^{2}\omega^{2}}{r}.
\end{equation}

Now using Kepler's third law ($GM=a^2 \omega^3$) the expression of amplitude can be written as
\begin{equation}
 h=\frac{G^{\frac{5}{3}}\mu M^{\frac{2}{3}} \omega^{\frac{2}{3}}}{c^{4} r}.  
\end{equation}

Moreover, the GW luminosity which may be defined as the total rate of energy loss at a distance $r$ related to GW over a spherical surface. Thus by definition luminosity can be given by~\cite{Bejger}
\begin{equation}
\textit{L}=\frac{dE_{GW}}{dt}.
\end{equation}

Using dimensional analysis we can write
\begin{equation}
\frac{dE_{GW}}{dt} \propto \frac{Gh^{2}\omega^{2}}{c^{5}}.
\end{equation}

Putting $h\propto \mu a^{2}\omega^{2}$ in the above equation, we get
\begin{equation}
\frac{dE_{GW}}{dt} \propto \frac{G\mu^{2} a^{4} \omega^{6}}{c^{5}}.
\end{equation} 

Now as we know that as a binary system revolves around each other looses energy. This loss in orbital energy supplies the necessary energy for a GW to leave the system. Total orbital energy for a binary system is given by the equation
\begin{equation}
 E_{orb}=-\frac{Gm_{1}m_{2}}{2a}.  
\end{equation}

Thus rate of loss of orbital energy can be given by
\begin{equation}
\frac{dE_{orb}}{dt}=\frac{Gm_{1}m_{2}\dot{a}}{2a^{2}}. 
\end{equation}

The above expression is equivalent to -$\frac{dE_{GW}}{dt}$ where $E_{GW}$ is the GW energy. Again by using Kepler's third law and obtaining the derivative of semi-major axis $a$, we finally can determine an expression for the growth of orbital frequency with time due to emission of GW. This can be provided as
\begin{equation}
\dot{\omega}=\frac{96G^{\frac{5}{3}}\omega^{\frac{11}{3}} M_ch^{\frac{5}{3}}}{5c^{5}}. 
\end{equation}

 The amplitude $h$ of GW also grows proportionally with the orbital frequency. This dual growth of frequency and amplitude is known as the chirp. The chirp mass is generally a function of the component masses, given by
\begin{equation}
  M_{ch}=\frac{(m_{1}m_{2})^{\frac{3}{5}}}{(m_{1}+m_{2})^{\frac{1}{5}} }.~\label{chirp}
\end{equation}

Orbital frequency of the system is directly related to the GW frequency $f_{GW}$ which is double that of the orbital frequency~\cite{Bejger}
\begin{equation}
f_{GW}=\frac{\omega}{\pi}.
\end{equation}

Therefore, a detector can directly measure the chirp mass by simply measuring the growth of $f_{GW}$ with time. Thus, by virtue of Eq. \ref{chirp}, the chirp mass in terms of the frequency $f_{GW}$ can be given by the following relation~\cite{Blanchet,Abbott1}
\begin{equation}
  \textit{M}=\frac{5c^{3}}{96G}\left(\pi^{-\frac{8}{3}}f_{GW}^{-\frac{11}{3}}\dot{f}_{GW}\right)^{\frac{3}{5}}.  
\end{equation}

Other information such as distance to source $r$, can be obtained using the time evolution of $f_{GW}$ and amplitude $h$. The GW signal has to be extracted by a detector from the noisy data. This is done by matched filtering techniques which we will discuss in the later Sections. 

\begin{figure}[h]
    \centering
    \includegraphics[width=10cm]{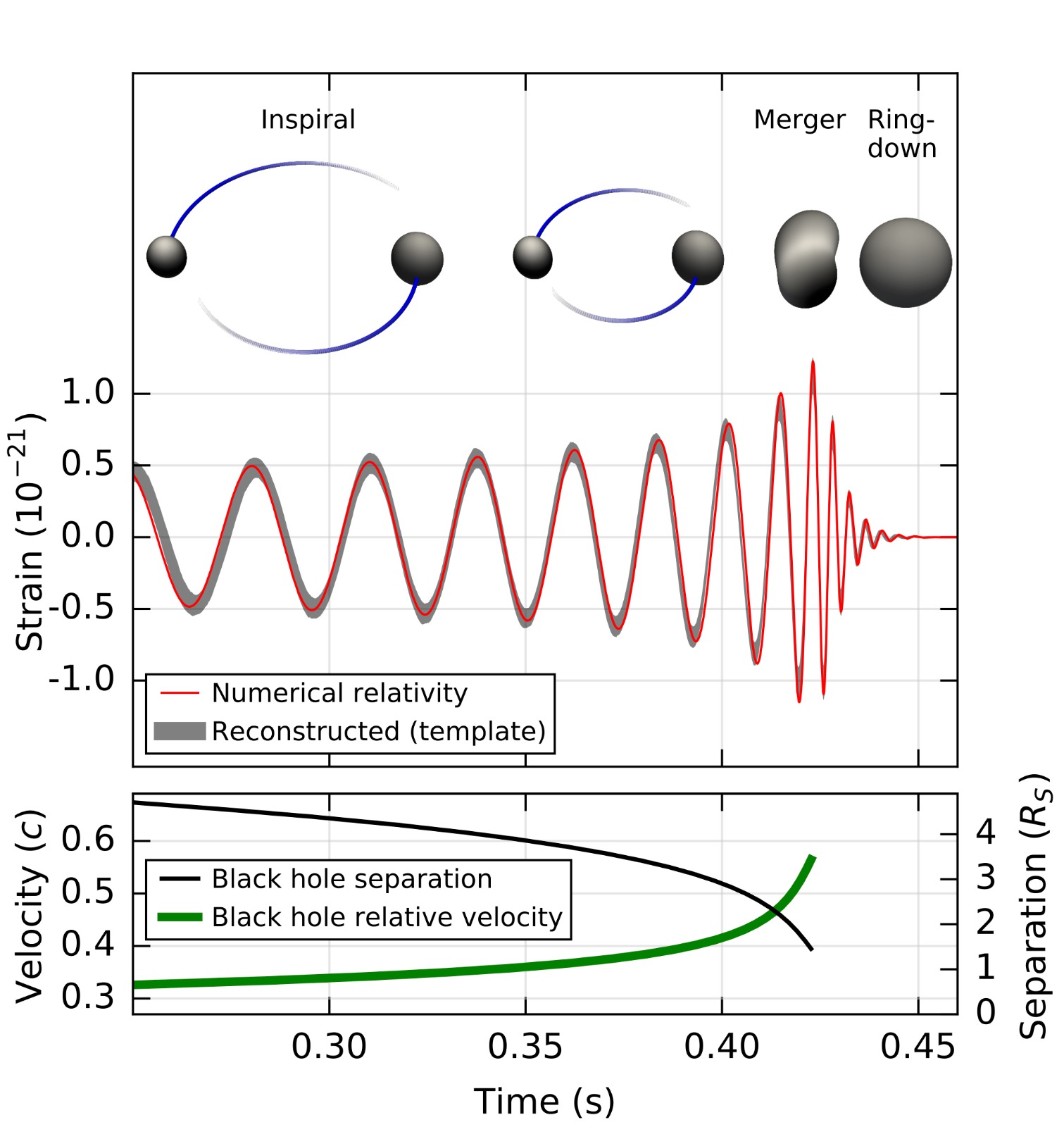}
    \caption{The first detection of GW150914 and its attributes. [Image Credit:  B.P. Abbott et al., PRL {\bf 116}, 061102 (2016)]}
    \label{fig2}
\end{figure}

\begin{figure}[h]
    \centering
    \includegraphics[width=12cm]{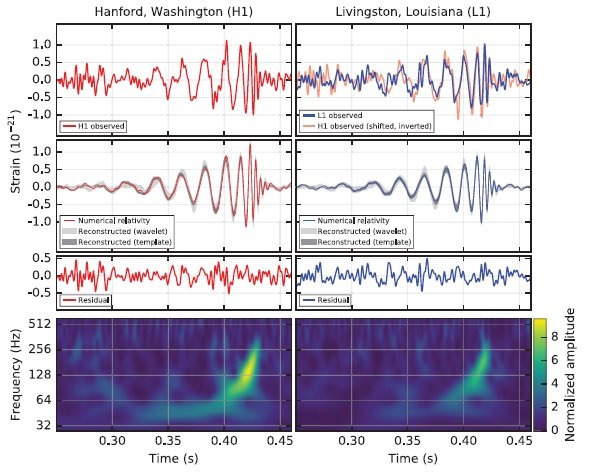}
    \caption{The signal from the gravitational wave event from GW150914, as detected by the LIGO Hanford (H1) and  Livingston (L1), respectively. [Image Credit: F.J. Raab and D.H. Reitze, Curr. Sci. 113, 657 (2017)]}
    \label{fig3}
\end{figure}

 \section{Detection of gravitational waves}
 
 \begin{figure}[h]
    \centering
    \includegraphics[width=12cm]{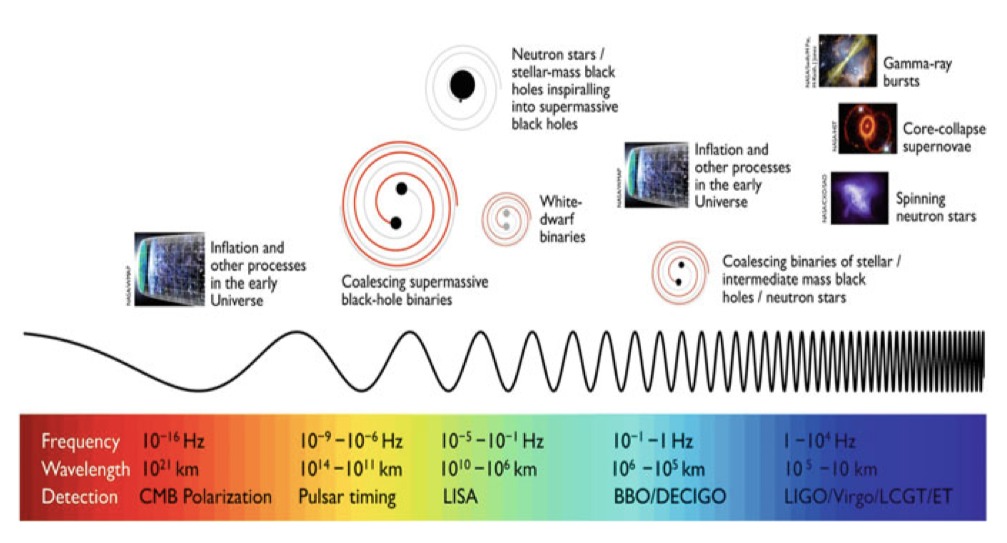}
    \caption{A schematic diagram of gravitational wave frequencies. [Image Credit: {\it Gravitational Waves: A New Window to the Universe}, A. Kembhavi and P. Khare, Springer Singapore (2020)]}
    \label{fig4}
\end{figure}

In 2015, the first GW produced by a binary black hole merger was detected by LIGO~(vide Figs. \ref{fig2} and \ref{fig3}). This led to increasing searches for GW in different frequency bands~\cite{Abbott1}. This can be classified on the basis of its frequency bands. Moreover, these frequency ranges are instrumental in deciding which detector can be used for a specific type of signal. The frequency classification of GW (Fig \ref{fig4}) are as follows~\cite{Wei}:
\\
\\
1) Ultra high frequency band (above 1 THz): Optical resonators and Terahertz resonators are basically used in detection of this type of GW.
\\
2) Very high frequency band (100 kHz–1 THz): Detection involves Optical interferometers, Gaussian beam detectors and Microwave resonator/wave guide detectors.
\\
3) High frequency band (10 Hz–100 kHz): Laser interferometric ground detectors and low temperature resonators are used for detection in this case.
\\
4) Middle frequency band (0.1 Hz–10 Hz): Space interferometric detectors are basically used for detection. These detectors are of short arm length(100 km-100000 km).
\\
5) Low frequency band (100 nHz–0.1 Hz): Laser interferometer space detector are primarily being used for detection in this case.
\\
6) Very low frequency band (300 pHz–100 nHz):Pulsar timing arrays are used for detection in this band.
\\
7) Ultra low frequency band (10 fHz–300 pHz): Detection in this range involves investigating quasar positions and proper motions.
\\
8) Extremely low (Hubble) frequency band(1 aHz–10 fHz): Cosmic microwave background experiments are used for detection in this range.
\\
9) Beyond Hubble frequency band (below 1 aHz): Direct verification is tough to find. Only indirect verification can be obtained using the inflationary cosmological models.

\subsection{Ground based observatories}

\begin{figure}[h]
    \centering
    \includegraphics[width=8cm]{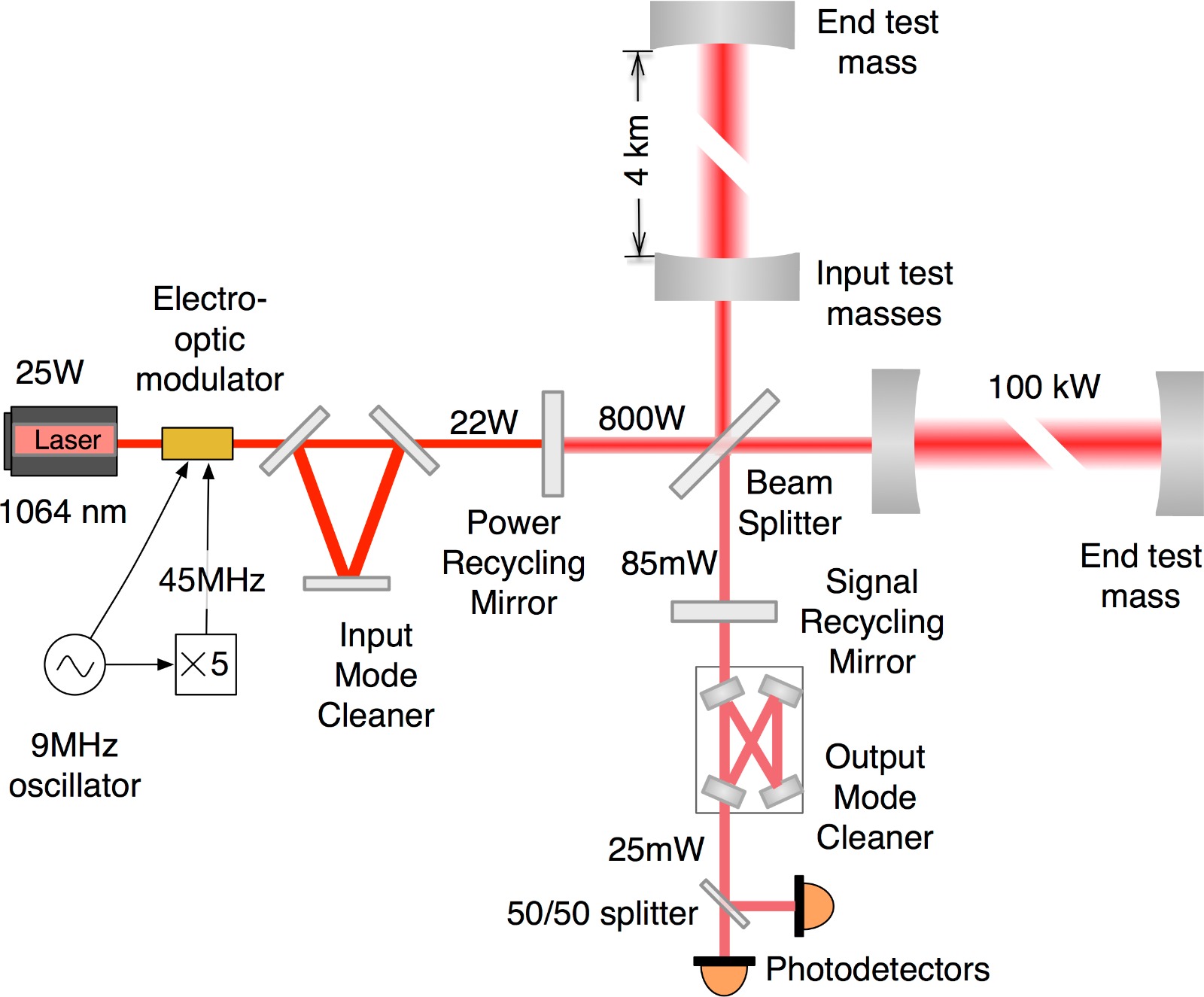}
    \caption{A schematic diagram of the Advanced LIGO interferometer. [Image Credit: D.V. Martynov et al., Phys. Rev. D 93, 112004 (2016)]}
    \label{fig5}
\end{figure}

\begin{figure}[h]
    \centering
    \includegraphics[width=12cm]{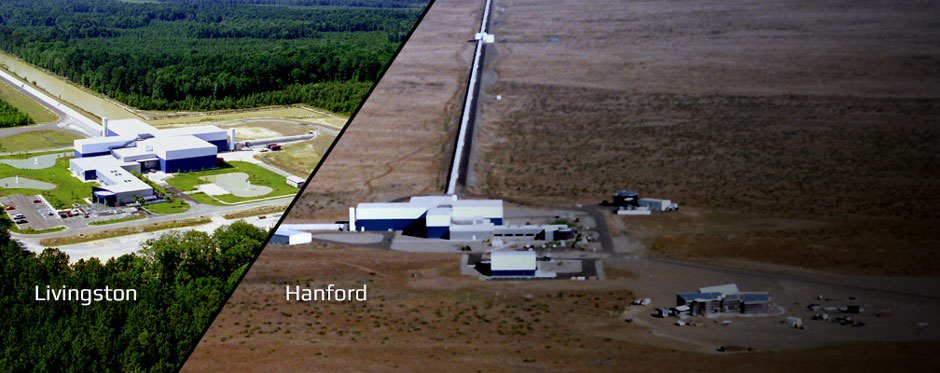}
    \caption{The aerial views of the LIGO observatory at Livingston and Hanford, respectively. [Image Credit: https://www.ligo.caltech.edu/]}
    \label{fig6}
\end{figure}

The first success in detection came from ground based observatories. Some of the ground based observatories for GW detection are LIGO, VIRGO and KAGRA. LIGO is located in the USA, whereas VIRGO is in ITALY and KAGRA is located in Japan. Typical frequency range for ground based observatories are 10 Hz-10 kHz.

In this Section, we will focus on LIGO which is one of the most prominent ground based GW observatory~(vide Figs. \ref{fig5} and \ref{fig6}). The primary goal for this observatory is to identify gravitational wave signals. Detectors used by LIGO are basically advanced Michelson interferometer. The interferometer used in LIGO has arms 4 km long which is approximately 360 times more than the  interferometer used in Michelson-Morley experiment. LIGO can sense the gravitational wave signal when it moves through both arms of the interferometer, as the arm lengths ($L$) get modified. Mathematically, this can be given by the following relation~\cite{Abbott1}
\begin{equation}
    \Delta L(t) \propto h(t) L,
\end{equation}
where the quantity $h(t)$ is known as the GW strain amplitude. From the above expression it can be inferred that the strain is proportional to length of the detector which is the primary reason for making the detectors substantially long. An equivalent amount of optical signal which is produced because of the altered arm length, thereby changing the phase difference between two light fields travelling towards the beam-splitter, gets transmitted to the photo-detector. Thus length of the arms are critical to the sensitivity of the detector. Moreover, to increase the effect of gravitational wave on the light phase a resonant optical cavity is used. The resonant optical cavity produced by two test mass mirrors in the LIGO set up can increase the effect of GW on the light phase by a huge factor. Numerically speaking, this factor is generally around 300~\cite{Rowan}.

Another factor responsible for detector sensitivity is the power-recycling mirror with fractional transmissivity, which increases the input laser power. At the output, a signal-recycling mirror optimizes the received gravitational wave signal by increasing the arm cavity bandwidth~\cite{Meers}. The kind of laser used in LIGO is a Nd:YAG laser which has fixed parameters such as its amplitude, frequency etc~\cite{Kwee}. Primarily, these advanced interferometric applications are instrumental in converting the developed strain to meaningful output signal. This also means reducing different kind of noise such as instrumental noises to improve the signal to noise ratio. In later sections we delve deeper into different kinds of noises and their statistics which one might come across while using these detectors.

\subsection{Space borne gravitational wave detectors}

Primary reason for developing a space based GW detector is to detect the GW signals that might have been generated due to supermassive black holes having masses in the range of $10^3-10^6 M_\odot$ where $M_\odot$ is the solar mass. Also, in terms of frequency, we try to detect GW signals having a frequency in the range of 0.1 mHz-1 Hz. Moreover, there might be stochastic GW signals as well which would have frequencies that are lesser than those detected by ground based observatories. In case of ground based observatories, they are less likely to detect a gravitational wave signal in lower frequency as its sensitivity gets limited due to the presence of seismic noise~\cite{Hough}. We can thus make use of the laser interferometry in space to successfully detect these signals. We can use similar procedures as a ground based observatory, i.e. comparing test mass distances using laser interferometry which will be inside different spacecrafts launched into the orbits~\cite{Pitkin}. 

\begin{figure}[h]
    \centering
    \includegraphics[width=10cm]{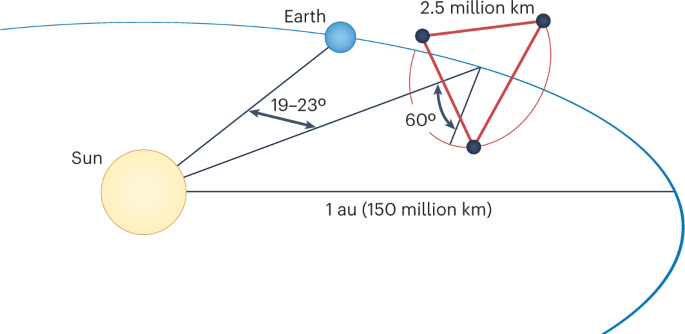}
    \caption{The orbiting LISA around the sun. [Image Credit: NASA, An Astro2020 White Paper -- The Laser Interferometer Space
Antenna]}
    \label{fig7}
\end{figure}

The most promising concept of detecting  GW in space can be attributed to LISA (Laser Interferometer Space Antenna)~[Fig. \ref{fig7}]. This has been a joint study by NASA and ESA~\cite{Stefano}. LISA would consist three different spacecraft which is scheduled to get launched by the year 2030. Once the spacecrafts reach their intended orbits which would be slightly tilted ~\cite{Jennrich} with respect to the ecliptic, at an angle of $60^{\circ}$ ~\cite{Pitkin}. These spacecrafts would then form an equilateral triangle where the distance between them would be $2.5 \times 10^{6}$ km~\cite{Gudrun}. Frequencies between 0.1 mHz to 1Hz is the working frequency range for LISA. This range enables LISA to detect  GW from a large number of compact binaries  with a larger period with which it orbits each other. This is in contrast to the detection done by the ground based observatories which deals with binary sources with much lesser period as it works in the frequency ranging from few Hz to kHz.Moreover, unlike ground-based observatories, with LISA it might be possible to detect  GW signals from white dwarfs as well.  However, as LISA would detect smaller frequencies it might end up detecting a large number of compact binaries which have frequencies in that range and hence causing superposed signals which is also known as confusion signal. Most of these detection though should have a high SNR and hence can be analysed separately~\cite{Gudrun}.

Apart from measuring  GW from compact binaries LISA would also be able to detect signals from binary black hole with masses equivalent to the stellar mass. Though ground based observatories has already detected signals from binary black holes still it could only detect that once their orbital distance decreases and frequency increases, but in case of LISA it would be able to sense the signal much before the merger phase. The working principle of LISA is more or less similar to the ground based observatories as LISA also would work in the principle of laser interferometry and the underlying concept as discussed in the case of ground based observatories remains the same here as well.

 Apart from LISA which seems to be the most promising concept for a space borne  GW detector, there are few other concepts such as ASTROD, ASTROD-GW (Astrodynamical Space Test of Relativity using Optical Devices), DECIGO (DECi-Hertz interferometer Gravitational Wave Observatory), BBO (Big Bang observatory). Working frequency range for ASTROD and ASTROD-GW should be similar to LISA. Though it would have a larger arm length than LISA (52 times longer than LISA) its detection sensitivity also should increase by a factor of 52. The set up has three spacecrafts which range interferometrically with each other and is at an arm length of 260 million km. The spacecrafts are in a formation resembling an equilateral triangle~\cite{Ni}. In case of DECIGO, which is a Japanese project, the frequency range is just in between the LISA and the ground based observatories( 0.1 Hz - 10 Hz)~\cite{Kawamura}. DECIGO would operate in a similar configuration as the LISA with three drag-free spacecrafts with the only exception of having a much shorter arm length than LISA at 1000 km only. 

BBO is another concept designed by the USA which would have similar working frequency range as the DECIGO. This would mostly focus in the detection of stochastic  GW background which would be a great way to get insights into the early universe and also in high precision cosmology~\cite{Cutler}. The configuration of BBO would be larger than LISA as it would contain three set of LISA like system with three spacecrafts each. With respect to its sensitivity it would be designed to be 2-3 times more sensitive than DECIGO~\cite{Pitkin}. 

\section{Inflationary gravitational waves: An overview}

In our previous section we discussed about the detection of stochastic  GW background. However the oldest relic of our universe, should come from primordial GW. To understand the generation of primordial GW we will include a brief insight into inflation and inflationary or primordial GW.  

Inflation may be explained as a process of accelerated expansion which started during initial moments of Big Bang~\cite{Albrecht}. Inflation is the main driving force behind our universe progressing towards a flat space geometry and overall homogeneity. However, quantum fluctuations from this period resulted in formation of galaxies, galaxy clusters and temperature anisotropies of the CMB (Cosmic Microwave Background)~\cite{Guth}. As we know that GW is generated due to spacetime perturbations, similarly in this case also inflationary GWs are generated as a result of quantum fluctuations of spacetime geometry~\cite{Weitou}.

Due to the primordial GW there must be imprinted tensor perturbation Microwave Background Radiation anisotropy. Several missions have been undertaken to analyse the tensor perturbation. One of them is WMAP (Wilkinson Microwave Anisotropy Probe), though 5 year data from the mission failed to discover any tensor perturbations~\cite{Weitou}. Several other mission such as PLANCK mission has been launched which would provide more accurate observational data for CMB polarization than WMAP~\cite{Planck}. Apart from this B-Pol~\cite{Bernardis}, EPIC~\cite{Bock} and Lite-BIRD~\cite{Hazumi} are some of the missions planned which should improve the sensitivity further. If we focus on the tensor modes in a CMB, it clearly points to two different causes, first one being the primordial gravitational waves~\cite{Smith} and other one being the interacting pseudo-scalar photos within the propagating CMB. Of course these possibilities can be differentiated based on detailed models and more accurate observations. If presence of primordial  GW gets confirmed then we can expect that a lot of insights on inflationary physics in future might just come from direct detection of primordial gravitational waves~\cite{Weitou}.

 Just to get a sense of the GW amplitudes for this specific type of GW, we can use the Friedmann equation which relates the expansion rate ($H_{in}$) with vacuum energy density $V$. The equation is given by
 \begin{equation}
  H_{in}= \frac{8\pi V}{3m_{pl}^{2}}, 
 \end{equation}
where $m_{pl}$ is the Planck's mass. 

Again, the vacuum energy density is proportional to the fourth power of Energy scale $E_{infl}$. This energy $E_{infl}$ is assumed to be the energy of the physical phenomenon responsible for the inflation. Therefore, it may be concluded that inflationary GW amplitude would depend on a particular inflationary model. As the models change the amplitude also differs~\cite{Zhang}. In conclusion, it can be predicted that depending on increasing accuracy and sensitivities of future detectors, detecting inflationary GWs remain a bright possibility.

\section{Highlights on data analysis by matched filtering}

Gravitational wave signals are generally very weak. Identifying GW therefore is not easy. In this paper, we focus on the detection and analysis techniques used by the interferometric observatories. In case of a GW signal, the extraction of the signal from the associated noise becomes a challenge. This can be done using matched filtering techniques (a statistical technique to be discussed in later section) which employ python or Matlab codes. To start a discussion on detection of gravitational wave signals we would start by enlisting different kinds of noises and their statistical nature.

\subsubsection{Random noise}
Instrumental noise can be categorised as a random noise. These type of noise can be defined as a series of random variables. Let us consider a random process which is a function of time and is given by $x(t)$. Now let us consider a probability density function $p_{x}$ for a definite value of $x$ at any time $t$. Therefore in this case the expectation value for this particular $x$ at time $t$ may be defined by the following  expression which is given by:     
\begin{equation}
\langle x \rangle:=\int xp_{x}(x) dx.\label{eq27}
\end{equation}
        
For a random process or noise if the statistical properties remains constant with time, then that can be defined as stationery random process. In this review, we discuss only the stationary and Gaussian random processes~\cite{Creighton}.

\subsubsection{Power Spectrum} 

Let us consider a signal x(t) with a zero mean. The power in the signal can be calculated by the following equation, 
\begin{equation}
\langle x^{2}(t) \rangle=\lim_{T\to\infty} \frac{1}{T} \int_{-T/2}^{T/2} x^{2}(t) dt.
\end{equation}

Considering $x(t)$ as stationary and $T$ is comparably large, the above equation is equal to that of the expectation value $\langle x^{2} \rangle$.
For a signal which is defined as
\begin{equation}
x_T(t)=\begin{cases}
x(t),~\text{if} -\frac{T}{2}<t<\frac{T}{2} \\
0\\ \text{otherwise}
\end{cases}
\end{equation}
     
Then the power spectral density for a process $x(t)$ may be defined as; 
\begin{equation}
\int_{0}^{\infty} S_x(f) dx.
\end{equation}
     
The above result is obtained after we apply Parseval's theorem to convert Eq. (\ref{eq27}) into frequency domain from time domain. Thus $S_x(f)$ is given by
\begin{equation}
S_x(f)= \lim_{T\to\infty} \frac{2}{T} \int_{0}^{\infty}| x^{2}(f)| df.
\end{equation}
     
The power spectral density $S_f(x)$ may be written as 
\begin{equation}
S_x(f)=\lim_{T\to\infty}\frac{2}{T} \left|\int_{-T/2}^{T/2} x(t)\exp(-2\pi\textit{i}ft) dt \right|^2.
\end{equation}

In the case of a stationary process, the power spectral density exceeds the auto-correlation function which is given by
\begin{equation}
    R_x(\tau)=\langle x(t) x(t+\tau)\rangle.
\end{equation}

We can also define the power spectral density in terms of the auto-correlation function  by the following equation
\begin{equation}
  S_x(f)=\int_{-\infty}^{\infty} R_x(\tau)\exp(-2\pi\textit{i}f\tau) d\tau. 
\end{equation}

Another popular expression for power spectral density in terms of the Fourier transforms of the time series can be given by the following equation
\begin{equation}
\langle \Tilde{x}^*(f') \Tilde{x}(f) \rangle=1/2(S_x(f)\delta(f-f')),
\end{equation}
where $\delta(f-f')$ is a Dirac delta function, $\Tilde{x}(f)$ is the Fourier transform of $x(t)$ and $\Tilde{x}^*(f')$ being its conjugate.

\subsubsection{Shot noise}

A shot noise or Poisson noise, may be defined as a random process which has a random number of pulses at random arrival times. Mathematically, this can be expressed as 
\begin{equation}
x(t)=\Sigma F(t-t_j),
\end{equation}
where $t_j$ is the random arrival time of $N$ number of pulses. The function $F(t)$ describes the pulse shapes. Thus, the average number of pulses arriving per unit time is given by, $R=N/T$ where T is the time period in which pulses arrive~\cite{Creighton}.
     
In this case the power spectral density is given by
\begin{equation}
S_x(f)=2R|F(f)|^{2},
\end{equation}
where we have assumed that $\langle \Tilde{x}(f) \rangle=0$.

For a broad-band spectrum and narrow pulse, we can assume $F(t)\approx \sigma \delta(t)$, thereby getting an expression for power spectral density in terms of an amplitude $\sigma$ which is given by
\begin{equation}
 S_x(f)=2R\sigma^2.   
\end{equation}

This expression does not depend on frequency and hence can be identified as a white noise~\cite{Creighton}.
      
\subsubsection{Gaussian noise}

Let us consider a noise with a time series $x(t)$ for total time $T$. This time series is sampled at regular interval, $\Delta t$ to produce $N$ samples. Each sample can be considered as independent and the probability distribution for all the samples may be given jointly as
\begin{equation}
p_x({x_j})=\left(\frac{1}{\sqrt{2\pi}\sigma}\right)^{N} e^{ -\frac{1}{2}\sigma^2 \sum x_j^2 }.
\end{equation}
      
This type of noise is known as Gaussian noise where it is assumed that $\sigma^2$ has both zero mean and variance. The power spectral density may be given as
\begin{equation}
S_x(f)= 2\sigma^2\Delta t,
\end{equation}
where $\Delta t$ is very small that is tending to zero. 

Noting the expression for power spectral density it may be inferred that as it does not depend on frequency, this can be considered to be a white noise. Moreover, the samples which are taken at random intervals are also independent of frequency, so Gaussian noise may be considered as a white noise. However, Gaussian noise might be a coloured noise as well and probability density for both types of Gaussian noise can be expressed by a single equation. This relation is given by~\cite{Creighton}
\begin{equation}
p_x[x(t)]\propto \exp\left[\frac{<x,x>}{2}\right],
\end{equation}
where $<x,x>$ is the inner product and $x(t)$ is a time series.

\section{Optimal detection statistics for a gravitational wave detection}

The main aim of a detector is to detect the GW signals, therefore if we have precise knowledge about the noise processes and the signal, we can make use of an optimal detection statistics. This type of statistics basically gives the probability that whether a set of data contains the anticipated signal or not. To understand this mathematically let us consider a set of strain data $s(t)$, which is being recorded by the GW detector. If $n(t)$ be a noise random process and the anticipated GW signal which is denoted by $h(t)$. Now to compute the probability, we begin by defining two different kind of hypothesis, the first one being null hypothesis which is given by
   \begin{equation}
    \textit{H}_0:  s(t)=n(t).    
   \end{equation}    
   
The second kind is known as alternative hypothesis which is given by the relation
\begin{equation}
   \textit{H}_1:  s(t)=n(t)+h(t). 
\end{equation}
    
The first case, that is the null hypothesis denotes that all of the strain data is basically noise and does not contain any GW signal, whereas the second case represents the strain data as the sum of noise and GW signal. To differentiate these hypothesis, we can calculate the ratio of the probability of null hypothesis being true for a particular set of strain data and the probability of alternate hypothesis being true for that set of strain data. This can be done by calculating the odds ratio. 

Mathematically odds ratio is represented by the relation, as~\cite{Creighton}
\begin{equation}
O(H_1|s)=P(H_1|s):P(H_0|s).
\end{equation}

Again, odds ratio can be computed using Bayes's theorem.

\subsection{Bayes's theorem}

For an event $A$, let us consider its probability to be true as $P(A)$ and similarly $P(B)$ must be the probability of an event $B$ being true. If $P(A|B)$ and $P(B|A)$ be the conditional probability respectively of an event being true given the other one is true, then this can be related to the individual probability by the relation
\begin{equation}
 P(A|B)=\frac{P(A,B)}{P(B)}.  
\end{equation}

Similarly for $P(B|A)$ we can write a similar relation which is given by
\begin{equation}
 P(A|B)= \frac{P(A,B)}{P(A)}.   
\end{equation}

Now from these two equations, Bayes's theorem can be written as
\begin{equation}
P(B|A)=\frac{P(B)P(A|B)}{P(A)}.
\end{equation}

Here $P(B)$ and $P(A)$ is defined as the marginal probability of $B$ and $A$ being true respectively. $P(B|A)$ is the probability of $B$ being true when $A$ is true and $P(A|B)$ is the probability of $A$ being true when $B$ is true. Bayes's theorem can be represented in another format which is given by
\begin{equation}
P(B|A)=\frac{P(B)P(A|B)}{(P(A|B)P(B)+P(A|-B)P(-B))}.
\end{equation}

This can again be written in terms of the likelihood ratio by the equation
\begin{equation}
  P(B|A)=\frac{\lambda(B|A)}{(\lambda(B|A)+P(-B)/P(B))},  
\end{equation}
where $P(-B)$ is equal to $1-P(B)$ and $P(A|-B)$ defines the probability of the event $A$ being true when $B$ is false. The likelihood ratio, thus can be defined as,
\begin{equation}
\Lambda(B|A)=\frac{P(A|B)}{P(A|-B)}.
\end{equation}
      
\subsection{Matched Filter}\label{sec:mf}

\begin{figure}[h]
    \centering
    \includegraphics[width=12cm]{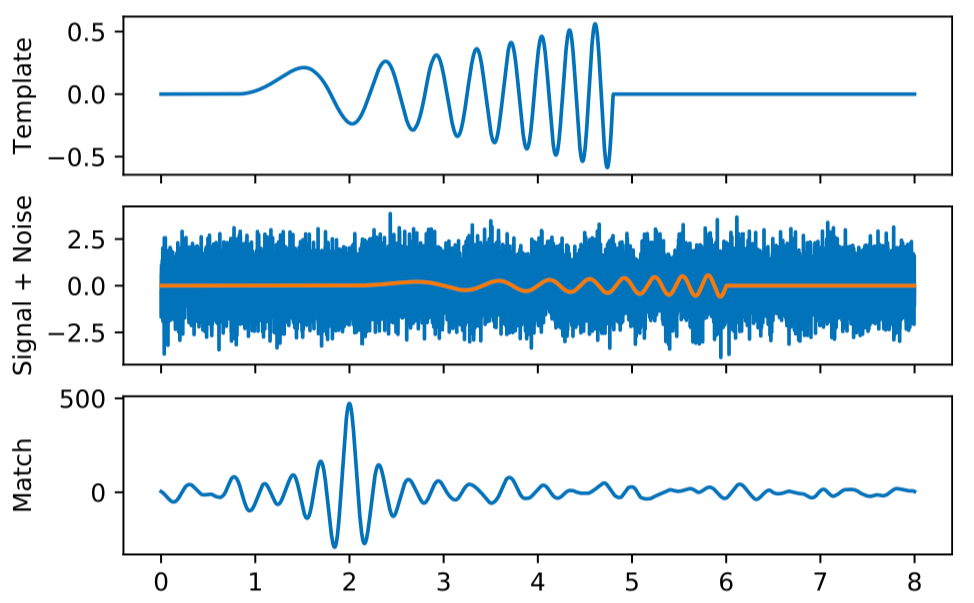}
    \caption{Depicting the basic operation of matched ﬁltering. [Image Credit: {\it General Relativity and Gravitational Waves: Essentials of Theory and Practice}, S. Dhurandhar and S. Mitra, Springer Nature Switzerland AG (2022)] }
    \label{fig8}
\end{figure}

Considering the null and alternative hypothesis and following Bayes's theorem we can calculate the likelihood ratio
\begin{equation}
\Lambda(H_1|s)=\frac{p(s|H_1)}{p(s|H_0)},
\end{equation}
where $p$ is the probability density and $H_0$ and $H_1$ is the null and alternative hypothesis respectively. 

Now if we assume the noise to be Gaussian the probability densities can be written for null hypothesis as
\begin{equation}
p(s|H_0)=p_n[s(t)]\propto \exp{\left[-\frac{<s,s>}{2}\right]}.
\end{equation}

Similarly for alternative hypothesis it may be written as
\begin{equation}
p(s|H_1)=p_n[s(t)-h(t)]\propto \exp{\left[-\frac{<s-h,s-h>}{2}\right]}
\end{equation}
      
Now the likelihood ratio in this case may be given by~\cite{Creighton}
\begin{equation}
\Lambda(H_1|s)=\exp{\left[\frac{<s,h>}{2}\right]}\exp{\left[-\frac{<h,h>}{2}\right]}. 
\end{equation}
      
 From the above relation we can infer that the likelihood ratio may be explained on the basis of a function which is monotonically increasing and depends on the inner product $<s,h>$. Thus this inner product can be regarded as the optimal detection statistic. The inner product $<s,h>$ known as the matched filter and plays the most important role in detection of GWs. This quantity is regarded as the matched filter since this can be used to deduce a correlation between the anticipated signal and the data in terms of signal to noise ratio~\cite{LIGO-2}.
      
Specifically in the cases of a low SNR, the signals generally remain hidden and matched filtering~(vide Fig. \ref{fig8}) is a handy tool to detect the GWs. In a matched filter the obtained data is matched over a large number of signal wave-forms also known as templates to find the best match. This process results in separation of the desired signals from the instrumental noise. Moreover, basic parameters of the source such as mass, spin, orbit orientation etc. can be estimated specifically by this technique~\cite{LIGO-2}.

\section{Deep Learning}

In the last decade deep learning (DL) usage has been grown very fast and became ubiquitous. It revolutionized the way the information are extracted hidden in the data in every sphere viz. health care, image recognition, natural language processing (NLP), language modelling, security, astronomy, etc.~\cite{ai2010,ai2014}. In this, artificial neural network (ANN) is the core of DL, in which Recurrent Neural Neworks (RNN) are suitable for the applications, such as NLP, language translation, speech recognition and image captioning. The second  important  DL algorithm is convolution neural network (CNN) which uses the concept of LeNet which can find key information contain in both image and time series data. Hence, it outperform for the image-related tasks, such as image recognition, object classification, pattern recognition and also deliver better performance in the classification of audio and signal data. To identify patterns within an image, CNN uses the principles of linear algebra, in particular matrix multiplication. Hence, arguably it is the most popular architecture in various fields, e.g. natural language processing, robotics, cybersecurity bio-informatics, malware detection, astronomy, etc. 

\begin{figure}
    \centering
    \includegraphics[scale=0.6]{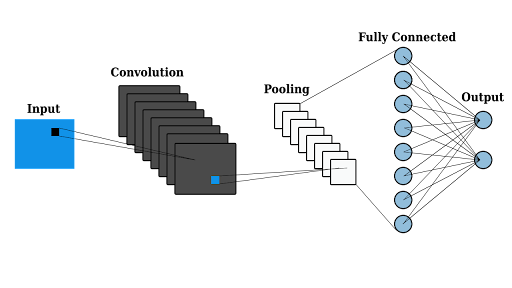}
    \caption{A typical Convolutional Neural Network (CNN)}
    \label{fig:cnn}
\end{figure}

The main advantage of CNN compared to its predecessors is that it automatically detects the important features without any human supervision. This is basically because of its three layers (convolution layer, pooling layer and fully connected layer) architecture (vide Fig. \ref{fig:cnn}), due to which the processing complexity increases, and in turn allows the CNN to successively identify larger portions and more complex features of an image until it finally identifies the object in its entirety. Hence, now it is a go-to model on every image related problem and time series data. Nevertheless, before deploying the model in real world problem, it is very important to know the performance of the designed model, and the metric used to measure the performance of the model is {\it accuracy} and {\it F1-score} and are given as
\begin{equation}
    \mbox{Accuracy} =  \frac{TP + TN}{TP + TN + FP + FN},
\end{equation}

\begin{equation}
\mbox{F1-Score} =  2 \times \frac{\text{Precision} \times \text{Recall}}{\text{Precision} + \text{Recall}},
\end{equation}
where \\
$\text{Precision} =  \frac{TP}{TP + FP}$, \\
$\text{Recall} =  \frac{TP}{TP + FN}$, \\
TP (True Positive) $\longrightarrow$ the number of positive class samples (i.e., GW signal)  that are correctly  
identify the presence of signal in the data,\\
TN (True Negative)  $\longrightarrow$  the number of negative class samples (i.e. noise) that are correctly says absence of signal in the data,\\
FP (False Positive)  $\longrightarrow$  the number of negative class samples (i.e. noise) that are incorrectly classified as signal,\\
FN (False Negative) $\longrightarrow$ the number of positive class samples (i.e. GW signal) that are incorrectly 
classified as noise.\\

The choice between this two metrics depends on the data-set used to train the model and how relevant is TP, TN, FN and FP. If data-set have an uneven class distribution  than F1-Score is recommended else accuracy. Also, accuracy is used when the TP and TN are more important while F1-score is used when the FN and FP are crucial.

\par Traditionally to  detect GW, matched filtering technique [Subsection \ref{sec:mf}]~\cite{match2004,match2014} are used in which banks of templates, which represent the expected signal waveform with all possible ranges of its parameters. In practice the bank of templates is matched to only a discrete set of signals from among the continuum of possible signals~\cite{Srivastava__2002}. Hence, because of the poor knowledge of the parameters space, this technique is computationally expensive to detect GW signals in the noisy data. However, the first direct detection of predicted gravitational waves (GW)~\cite{Einstein1916b} by the advanced Laser Interferometer Gravitational Wave Observatory (aLIGO) is done by matched filtering technique, and since then, almost hundred compact binary coalescences (CBCs) have been detected ~\cite{GW151226,GW170104,abbott2019binary,abbott2019gwtc,abbott2021gwtc,abbott2021gwtc1} using matched filtering. Understanding that the matched filtering is computationally costly in particular for the detection of continuous gravitational wave and the recent diverse effective application of DL, researcher started exploring different architectures  of DL~\cite{Sewak:2020:1546-1955:182} for the detection of GW efficiently. It's in nascent stage, however it has been observed that CNN can identify the GW signal in the noisy data, because it can process images and time-series data like audio stream, which are structurally similar to the strain produced by GW observatories. 

Therefore, it has been noted that majority of the work revolves on CNNs  to identify different types of GW signals~\cite{dreissigacker2020deep,gabbard,fan2019applying,PhysRevD.97.044039,bahaadini2017deep,krastev2020real,lin2020binary,li}, glitch classification~\cite{zevin2017gravity,coughlin2019classifying,razzano2018image} and parameter estimation~\cite{george2017deep,george2018deep,krastev2021detection} in the output of the GW detectors. In this, Sara et al.~\cite{bahaadini2017deep}, proposed a deep multi-view CNN for the classification of the non-Gaussian disturbances known as glitches. Their model can accommodate efficiently the various classes independently of the glitch duration, hence improve the overall classification accuracy  compared to traditional single view algorithms. In 2018, George et al.~\cite{PhysRevD.97.044039} using CNNs proposed a {\it Deep Filtering} method to detect GW signals in white noise time-series data and estimated the parameters of their sources in real time. Their analysis shows that that the {\it Deep Filtering} outperforms conventional machine learning techniques and gave similar performance to the most sensitive algorithms (matched filtering) used for the detection of GWs. Also their approach being several orders of magnitude faster, hence they claimed that that their approach can facilitate real-time searches of GW events in real LIGO data. Later, Gabbard et al.~\cite{gabbard} constructed a deep CNN network for the search of binary black hole GW signals. For the experimental analysis they used whitened time series GW strain to train and test the model on simulated binary black hole signals. Their analysis shows that the proposed network can classify signal from noise with sensitivity similar to matched-filtering technique.

\par Recently, Li et. al.~\cite{li} studied the optimization techniques and proposed a scheme based on CNN and wavelet packet decomposition method for the detection of GW. They decomposed the data using a wavelet packet, representing the GW signal and noise using the derived decomposition coefficients; and then determined GW existence event using CNN with a logistic regression output layer. They claimed that their proposed scheme detection sensitivity is higher compared to~\cite{gabbard}. However, the practical implications of their work are constrained by their choice of noise and the simplified damped sinusoidal as an analytical waveform model. Later, Krastev~\cite{krastev2020real}, proposed a method using CNN to identify transient GW signals from binary neutron star mergers in noisy time series representative of typical gravitational-wave detector data. They shown that a deep CNN trained on 100,000 data samples can detect the binary neutron star GW signals and distinguish them from noise and signals from merging black hole binaries. They also demonstrated that ANN can detect real-time GW signals emitted from the binary neutron star mergers. Also Lokesh et al.~\cite{lokesh_cascade} proposed and investigate a two-step cascaded classification model using CNN for the detection of GW signals emitting from the two sources, i.e. BBH and BNS.  In their proposed model, they first used the CNN to know whether in the noisy data stream, the GW signal (BBH/BNS) is present or not. While in the second step, they further applied CNN to know that the present GW signal in the stream is BBH or BNS. Their analysis shows that the proposed two-step cascaded classification model can detect not only the presence of the signal but also able to distinguish between the type of signal, i.e. whether the GW signal is from BBH or BNS. Hence,  we can say that CNN can be promising DL approach in future for the detection of different types of GW buried in the noise.

\section{Conclusion}
In this review paper we have discussed the theoretical concepts relating to the generation of GWs, sources and its detection techniques. We also have discussed about matched filter which is an optimal technique to detect gravitational waves at the output of a GW detector. Some of the very recent developments as detection techniques such as deep learning, especially CNN, also has been provided here. 

Study of GWs from some far flung celestial object can provide detailed information about the nature of its source and surroundings. Gravitational wave observation can therefore open new pathways to study the mass, spin, deformity and eccentricity of a heavenly system. However, eccentricity comes into play only in case of binary system and deformity is mainly in case of rotating objects. 

Finally, the most important inference drawn in terms of modern physics as well as astrophysics was that the detection of spacetime ripples, i.e. GWs from a black hole merger in 2015 by detectors like advanced LIGO and VIRGO, proved Einstein's GR is correct. As the GW has been detected firmly after a century of its prediction, therefore it has presently paved the path for new challenges in terms of growth of more modern observatories with better sensitivity and cutting edge technology. In this avenue of research following various pulsar timing array (PTA) experiments (such as NANOGrav, EPTA, PPTA, CPTA, including data from InPTA) has been already started to gather new information and knowlege on gravitational wave astronomy regarding which a bit we have mentioned in the Introduction~\cite{SRC2021,Guo2023,Vagnozzi2023,Abbott3,Abbott4,Khlopov2010,Belotsky2019}.

\section*{acknowledgments}

SR and SKS are thankful for the support from the Inter-University Centre for Astronomy and Astrophysics (IUCAA), Pune, India for providing the working facilities and hospitality under the respective Associateship schemes. A part of this work was completed while RB was visiting IUCAA and the author gratefully acknowledges the warm hospitality and facilities there.

\end{document}